\documentclass{cupconf}
\usepackage{natbib,graphicx}

\def\lesssim{\mathrel{\hbox{\rlap{\hbox{\lower4pt\hbox{$\sim$}}}\hbox{$<$}}}}
\def\gtrsim{\mathrel{\hbox{\rlap{\hbox{\lower4pt\hbox{$\sim$}}}\hbox{$>$}}}}
\let\la=\lesssim
\let\ga=\gtrsim
\newcommand{\gt}{$>$}
\newcommand{\msun}{M_{\odot}}
\newcommand{\lsun}{L_{\odot}}
\newcommand{\rsun}{R_{\odot}}
\newcommand{\avir}{\alpha_{\rm vir}}
\newcommand{\calm}{\mathcal{M}}

\newcommand{\apj}{ApJ}
\newcommand{\apjs}{ApJS}
\newcommand{\apjl}{ApJ}
\newcommand{\aap}{A\&A}
\newcommand{\mnras}{MNRAS}
\newcommand{\pasp}{PASP}
\newcommand{\nat}{Nature}
\newcommand{\araa}{ARA\&A}
\newcommand{\aj}{AJ}

\title[Star Formation by Core Collapse]{High Mass Star Formation by Gravitational Collapse of Massive Cores}

\author[M.~R. Krumholz]{
M\ls A\ls R\ls K\ns R.\ns K\ls R\ls U\ls M\ls H\ls O\ls L\ls Z\ls$^1$\footnote{Hubble Fellow}}

\affiliation{$^1$ Department of Astrophysical Sciences, Princeton University, Princeton, NJ 08544-1001}

\begin{document}

\maketitle

\begin{abstract}
The current generation of millimeter interferometers have revealed a population of compact ($r \lesssim 0.1$ pc), massive ($M \sim 100$ $\msun$) gas cores that are the likely progenitors of massive stars. I review models for the evolution of these objects from the observed massive core phase through collapse and into massive star formation, with particular attention to the least well-understood aspects of the problem: fragmentation during collapse, interactions of newborn stars with the gas outside their parent core, and the effects of radiation pressure feedback. Through a combination of observation, analytic argument, and numerical simulation, I develop a model for massive star formation by gravitational collapse in which massive cores collapse to produce single stars or (more commonly) small-multiple systems, and these stars do not gain significant mass from outside their parent core by accretion of either gas or other stars. Collapse is only very slightly inhibited by feedback from the massive star, thanks to beaming of the radiation by a combination of protostellar outflows and radiation-hydrodynamic instabilities. Based on these findings, I argue that many of the observed properties of young star clusters can be understood as direct translations of the properties of their gas phase progenitors. Finally, I discuss unsolved problems in the theory of massive star formation, and directions for future work on them.
\end{abstract}

\firstsection
\section{Introduction}

Massive star formation occurs in the densest, darkest parts of molecular clouds. These clumps of gas have masses of thousands of $\msun$, radii $\la 1$ pc, volume densities of $\sim 10^5$ H atoms cm$^{-3}$, column densities of $\sim 1$ g cm$^{-2}$, visual extinctions of hundreds, and velocity dispersions of several km s$^{-1}$. Observations often reveal indicators of massive star formation such as maser emission and infrared point sources within them, but the majority of their mass appears to be dark and cold. Due to their high extinctions and low temperatures, these regions are only accessible to observation through millimeter emission, either in molecular lines \citep[e.g.][]{plume97,shirley03,yonekura05,pillai06b} or dust continuum \citep[e.g.][]{carey00,mueller02}, or through infrared absorption \citep[e.g.][]{egan98,menten05,rathborne05,simon06,rathborne06a}. They are likely the progenitors of the rich clusters in which massive stars form.

In the last few years observations using millimeter interferometers to obtain still higher resolution have identified ``cores" within these dense clumps, objects small enough that they approach the stellar mass scale. Cores are distinguished by even higher volume densities than the massive clumps clumps around them, $10^6$ H cm$^{-3}$ or more, and smaller radii, $r \la 0.1$ pc.  Some show mid-IR point sources in their centers \citep[e.g.][]{pillai06a}, while other show no MIR emission or even MIR absorption, indicating that they are starless or contain only very low mass stars \citep[e.g.][]{sridharan05}. In some cases they show signs of molecular outflows but not MIR emission, indicating that the extremely massive core contains a very low mass protostar, and thus is near the onset of star formation \citep{beuther05b}.

The characteristic mass, size, and density of massive cores make them appealing candidates to be the progenitors of massive stars \citep[e.g.][]{garay05}. Moreover, as I discuss in more detail in \S~\ref{initial}, the observed mass and spatial distribution of protostellar cores is quite similar to that of stars in young clusters. If massive cores are the direct precursors of massive stars, then one can explain many of the properties of newborn clusters directly from the observed properties of their gas phase progenitors. The goal of this review is to construct a rough scenario based on this idea by following observed cores through collapse, fragmentation, accretion, and feedback, to the final formation of massive stars. In \S~\ref{initial} I briefly review observations of the properties of massive cores to provide initial conditions for this scenario. In \S~\ref{fragmentation}, \S~\ref{compaccretion}, and \S~\ref{feedback} I discuss three major questions about how these cores turn into stars: do they fragment into many stars or only a handful? Do the stars they form accrete significant mass from outside the parent core? Does feedback significantly inhibit accretion? Finally, in \S~\ref{future} I discuss some outstanding problems in the modeling of massive core evolution, and suggest directions for future work.

\section{Massive Cores: Initial Conditions for Massive Star Formation}
\label{initial}

We know disappointingly little about massive cores, despite great observational effort. Due to their small sizes and large distances, massive cores are only marginally resolved even in observations with the highest resolution telescopes available. Nonetheless, observations do allow us to determine some gross properties of individual massive cores, and of the massive core population as a whole. Observations show that cores are centrally concentrated, although the exact density profile is difficult to determine with interferometer measurements, and fairly round, with aspect ratios of roughly 2:1 or less. They are cold, $T\approx 10-40$ K, except near stellar sources, so their observed velocity dispersions of $\sim 1$ km s$^{-1}$ imply the presence of highly supersonic motions \citep[e.g.][]{reid05,beuther05b,beuther06a}. At the characteristic density of $\sim 10^6$ H cm$^{-3}$ found in these cores, the free-fall time is only $\sim 10^5$ yr, so the implied accretion rate when a massive core collapses is $10^{-4}-10^{-3}$ $\msun$ yr$^{-1}$. 

\citet{mckee03} propose a simple self-similar model of massive cores in which the core density and velocity dispersion are power law functions of radius such that at every radius turbulent motions provide enough ram pressure to marginally support the core against collapse. The central idea is that, at the high pressures found in massive star-forming regions massive cores must be supported by internal turbulent motions. While a self-similar spherical model is obviously a significant simplification of a turbulently-supported gas cloud, it provides a reasonably good fit to the available observations, and makes it possible to calculate quantities such as the time scale for star formation and the relationship between core mass, column density, pressure, and velocity dispersion. It also provides a good starting point for simulations and more detailed analytic work.

For the massive core population as a whole, we know somewhat more, and the observations bolster the idea that massive cores might really be the progenitors of massive stars. Several authors using different techniques and observing different regions find that the mass distribution of massive cores matches the stellar initial mass function, shifted to higher mass by a factor of a few, with a Salpeter slope of $\Gamma \approx -2.3$ at high masses and a flattening at lower masses \citep{beuther04b, reid05, reid06a, reid06b}. This extends earlier observational work indicating that in low and intermediate mass star-forming regions the core mass function resembles the stellar IMF \citep[e.g.][]{motte98,testi98,johnstone01,onishi02}, and suggests that the stellar IMF may be set simply by the mass distribution of prestellar cores, reduced by a factor of a few due to mass ejection by protostellar outflows \citep{matzner00}. Simulations and analytic arguments, can, in turn, explain the core mass distribution as arising naturally from the supersonic turbulence present in star-forming clumps \citep{padoan02,tilley04,li04}.

\citet{clark06} argue that the mass distribution of \textit{bound} cores in simulations does not have a Salpeter slope and thus is unlikely to be the origin of the stellar IMF. However, this misses a critical point: the Salpeter slope is only observed for stars significantly above the peak of the IMF. The full IMF is closer to a broken power law \citep{kroupa02} or a lognormal \citep{chabrier03}, with the break or peak at $\sim 0.5$ $\msun$. This is roughly the Jeans mass in star-forming clumps, and indeed the simulations of \citet{clark06} do show something like a lognormal distribution, with a peak at roughly the Jeans mass of their simulations. (The simulations are scale-free, since they include only hydrodynamics and gravity and use an isothermal equation of state.)

Furthermore, recent observations focusing on the spatial distribution of cores have shown that cores are mass segregated \citep{stanke06} in much the same manner as stars in very young clusters: the core mass function has the same lognormal or broken power law form everywhere in clump, with the exception that the most massive cores, those larger than a few $\msun$ in size, are found only in the very center. The stellar population of the ONC follows the same pattern \citep{hillenbrand98, huff06}, indicating that the observed mass segregation in stars may simply be an imprinting of the prestellar core spatial distribution. At least some of the mass segregation must be primordial rather than a result of dynamical evolution \citep{bonnell98b}, although recent evidence that cluster formation takes several crossing times \citep{tan06a} suggests that evolution may be important too. Nonetheless, it is quite suggestive that both the IMF and the spatial distribution of stars in a cluster seem to be explicable solely from the observed distribution of gas from which star clusters form. However, the origin of the mass segregation of cores is at present unknown.

\section{Fragmentation of Massive Cores}
\label{fragmentation}

It is only possible to explain the properties of stars in terms of the properties of cores if there is a more or less direct mapping from core mass to star mass. Such a mapping exists only if cores do not fragment too strongly, i.e. if massive cores typically produce a one or a few massive stars, rather than many low stars. Fragmentation to a few objects does not present a problem, since observationally-constructed mass functions are generally uncorrected for multiple systems, but fragmentation to many objects does.

One might expect massive cores to fragment because they contain many thermal Jeans masses of gas. At the densities of $\sim 10^6$ H cm$^{-3}$ and temperatures of $\sim 10$ K typical of massive cores, the Jeans mass is only $\sim \msun$, so one might expect a 50 $\msun$ core to form tens of stars. \citet{dobbs05} simulate the collapse and fragmentation of massive cores with initial conditions based on the \citet{mckee03} model, using a code that includes hydrodynamics and gravity. They try both isothermal and barotropic equations of state. (Barotropic here means that the gas is assumed to be isothermal at densities below some a critical density, chosen to be $10^{-14}$ g cm$^{-3}$ in the Dobbs \textit{et al.} simulations, and adiabatic at higher densities). Dobbs \textit{et al.}  find that the cores fragment, forming anywhere from a few to several tens of objects, depending on the assumed initial conditions and equation of state. In no case do their simulations from a massive star.

However, the \citet{dobbs05} calculation omits the critical effect of radiation feedback from the forming star. The high densities in massive cores produce high accretion rates, so that the first protostar to condense within a core will immediately produce a large accretion luminosity as the gas that falls onto it radiates away its potential energy. For a typical accretion rate of 
$10^{-4}$ $\msun$ yr$^{-1}$ at massive core densities, a $0.1$ $\msun$, $1$ $\rsun$ star releases approximately $300$ $\lsun$ of accretion power. Because the core is very optically thick, the radiation is trapped within it and heats the gas as it diffuses out. As a result, the densest, inner parts of the core where fragmentation is most likely to take place are subject to rapid heating, which suppresses fragmentation. Isothermal or barotropic approximations completely ignore this effect.

\begin{figure}
\begin{center}
\includegraphics{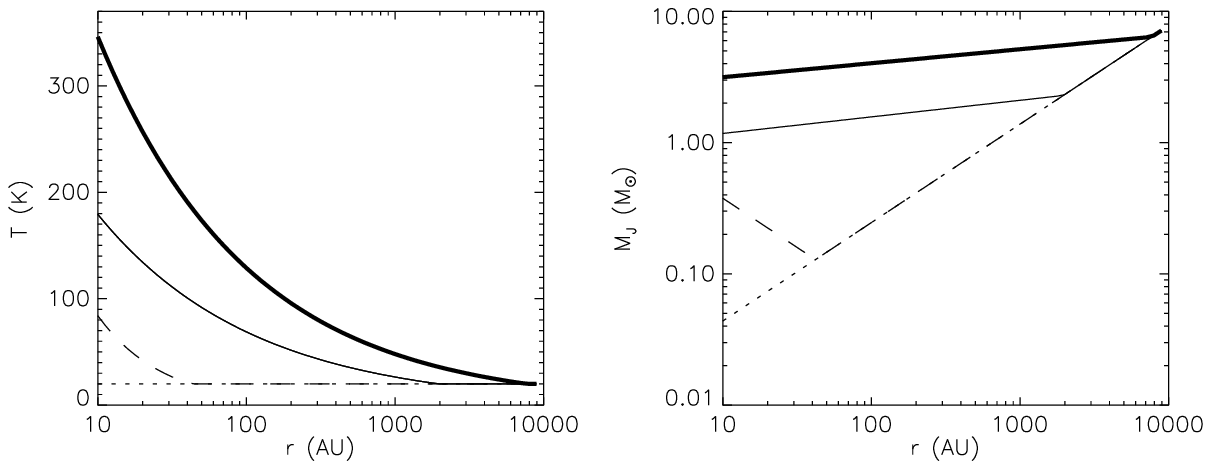}
\end{center}
\caption{
\label{radfrag}
Temperature (\textit{left panel}) and Jeans mass (\textit{right panel}) versus radius in a core with a mass of 50 $\msun$, a column density of 1 g cm$^{-2}$, and a density profile $\rho\propto r^{-1.5}$, taken from the models of \citet{krumholz06d}. The lines show the result from a radiative transfer calculation when the central protostar is 0.05 $\msun$ (\textit{thin solid line}) and when it is 0.8 $\msun$ (\textit{thick solid line}), and from using the barotropic approximation (\textit{dashed line}) or an isothermal equation of state (\textit{dotted line}). The Jeans mass is computed using the density and temperature at each radius, and is defined as $M_J = \frac{4}{3} \pi^{5/2} [k_B T/(G \mu)]^{3/2} \rho^{-1/2}$, where $\rho$ is the density, $T$ is the temperature, and $\mu=2.33m_H$ is the mass per particle for a gas of molecular hydrogen and helium in the standard interstellar abundance.
}
\end{figure}

\citet{krumholz06d} examines how feedback heating affects fragmentation in the context of a simple model of core accretion using a high accuracy analytic radiative transfer approximation \citep{chakrabarti05}. Figure \ref{radfrag} shows a sample result, the radial temperature profile and Jeans mass versus radius for a \citet{mckee03} core with a mass of 50 $\msun$ and a column density of 1 g cm$^{-2}$ accreting onto a protostar in its center. The figure compares the results using a radiative transfer approach to what one would find by neglecting radiative transfer and simply using a barotropic or isothermal equation of state. As the plot shows, both an isothermal equation of state and the barotropic approximation make order of magnitude errors in the temperature and Jeans mass.

One might worry whether this analytic treatment done in spherical symmetry applies to more realistic massive cores with complex density structures \citep[e.g.][]{bonnell06a}. The natural way to address this question is with radiation-hydrodynamic simulations of massive core evolution. Krumholz, McKee, \& Klein (2006, in preparation) simulate the collapse and fragmentation of massive cores to examine this effect. The simulations use an adaptive mesh refinement radiation code to solve the Euler equations of gas dynamics coupled to gray radiation transport and radiation pressure force in the flux limited diffusion approximation \citep{truelove98,klein99,howell03}. They use the adaptive mesh capability to guarantee that the local Jeans length is always resolved by at least 8 cells \citep{truelove97}, and that the radiation energy density changes by no more than 25\% per cell, so radiation gradients are well resolved. The code uses Eulerian sink particles to represent stars \citep{krumholz04}, and the sink particles are in turn coupled to a simple protostellar evolution model \citep{mckee03} which computes the instantaneous stellar luminosity, including the effects of accretion, Kelvin-Helmholtz contraction, deuterium burning, and hydrogen burning. This luminosity becomes a source term in the radiation equation. Further details on the code are given in \citet{krumholz05d}.

The simulations begin with cores following the model of \citet{mckee03}. The initial density profile is chosen with $\rho \propto r^{-1.5}$, to a maximum density of $\rho=10^{-14}$ g cm$^{-3}$, corresponding roughly to the density of the inner, thermally-supported zone of McKee \& Tan cores. The temperature is 20 K throughout the core. There are initial turbulent velocities chosen from a Gaussian random distribution \citep{dubinski95} with a power spectrum $P(k)\propto k^{-4} d^3 k$ over wavelengths ranging from the size of the core to the size of the inner thermal zone, subject to the constraint that the initial velocity field be divergence-free. The magnitude of the velocity field is normalized to give approximate hydrostatic balance on the largest scale \citep[equation 6 of][]{mckee03}. The simulations reach a maximum resolution of 10 AU.

\begin{figure}
\begin{center}
\includegraphics{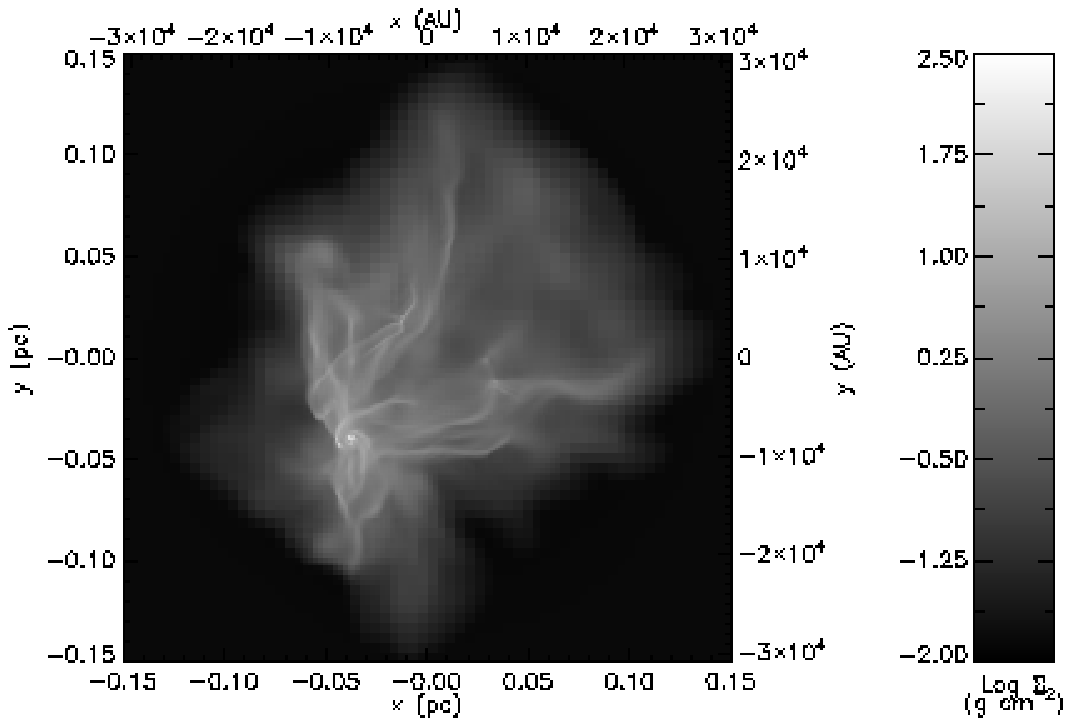}
\includegraphics{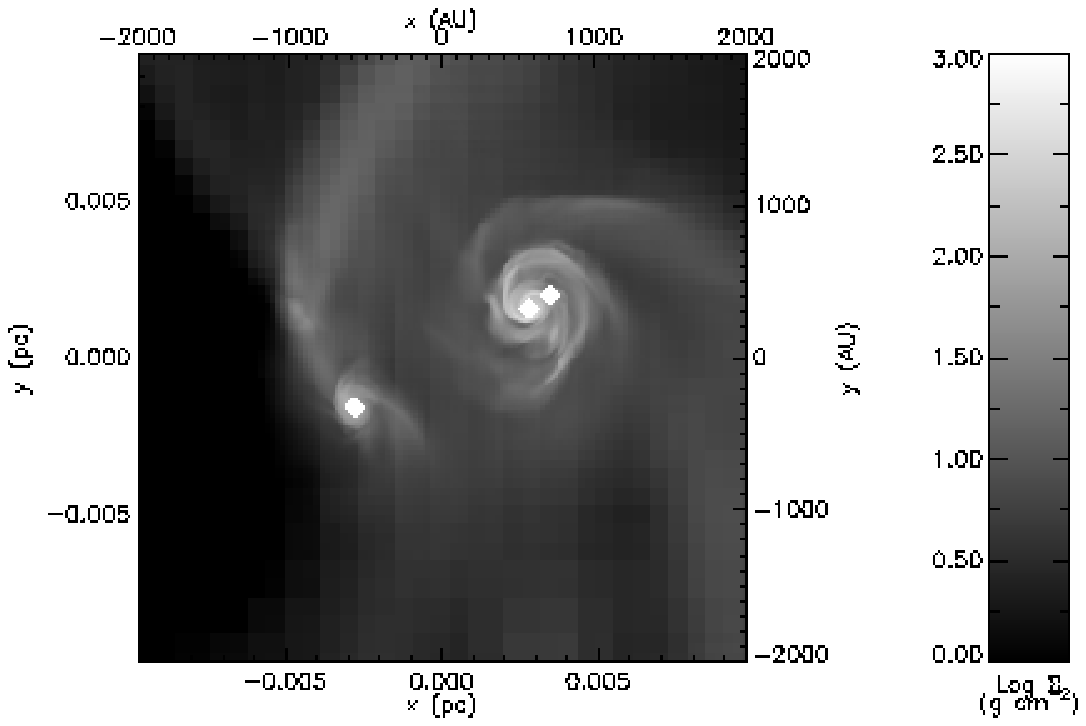}
\end{center}
\caption{
\label{columnfrag}
Column density of the entire core (\textit{upper panel}) and zoomed in on the forming stars (\textit{lower panel}) for a simulation of a $100$ $\msun$, 0.1 pc \citet{mckee03} core at a time of $2.0\times 10^4$ yr.  The positions of the stars are indicated by the diamonds. Their masses are, from left to right, $0.31$ $\msun$, $5.33$ $\msun$, and $0.16$ $\msun$.
}
\end{figure}

Figure \ref{columnfrag} shows the column density distribution in a simulation of a core with an initial mass of $100$ $\msun$ and radius of $0.1$ pc, $2.0\times 10^4$ yr (0.37 mean-density free-fall times) after the start of a simulation. The core is not forming many stars, it is forming a triple system. Moreover, it is a highly unequal triple: the masses of the three stars are $5.33$ $\msun$, $0.31$ $\msun$, and $0.16$ $\msun$, so the vast majority of the mass has gone into the most massive object, the one at the center of the large disk. There are no apparent signs of further fragmentation, so unless feedback disrupts this system, it seems destined to form a massive star incorporating a significant fraction of the initial core mass rather than dozens of small stars. The weak fragmentation we find from simulations provides strong support for the idea that the core mass function directly sets the stellar mass function.

\begin{figure}
\begin{center}
\includegraphics{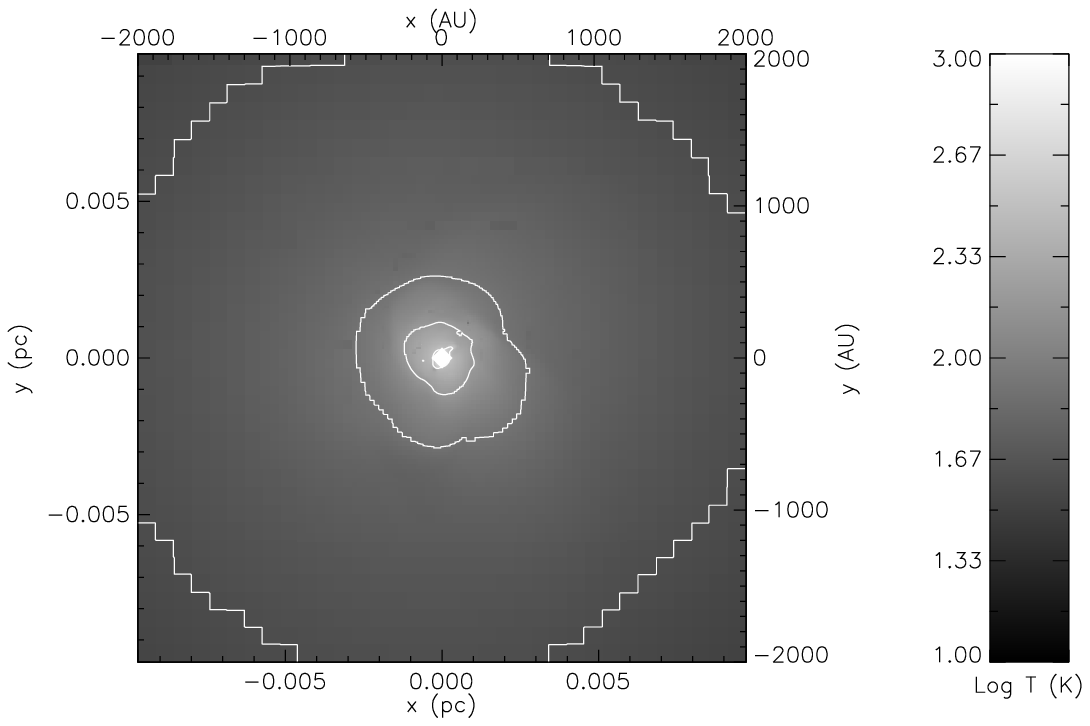}
\end{center}
\caption{
\label{tempfrag}
Temperature distribution in a 2D slice through a simulation of a $100$ $\msun$, 0.1 pc \citet{mckee03} core at a time of $1.6\times 10^4$ yr. The image is centered on the $3.83$ $\msun$ star, indicated by the diamond, the most massive present in the simulation at that time. The outermost contour corresponds to a temperature of 40 K, and each subsequent contour represents a factor of 2 increase in the temperature.
}
\end{figure}

To understand the origin of the weak fragmentation, it is helpful to examine the temperature distribution in the core. Figure \ref{tempfrag} shows the temperature distribution in the simulation at $t=1.6\times 10^4$ yr, when the central star is only $3.83$ $\msun$. At this point the star has not yet begun hydrogen burning, and the luminosity of a few thousand $\lsun$ is entirely due to accretion. This accretion power has doubled the initial temperature of the gas out to more than 2000 AU from the central star, and increased the temperature to more than 100 K  over a radius of many hundreds of AU. This heating strongly suppresses fragmentation in the densest gas, where it is most likely to occur. Of the two stars that do form in addition to the most massive, one does so at an initial separation of several thousand AU, far enough that it can condense, and the other does so inside the protostellar disk, where the high column density provides shielding against the stellar radiation and produces a lower temperature. In examining a movie of the simulation, one clearly sees many overdense clumps that look like they might collapse, but do not do so because they are bathed in the radiation field of the central star. Rather than forming stars of their own, they fall onto the central star and accrete.

Part of the reason suppression of fragmentation is effective is because, despite the complex density structure shown in Figure \ref{columnfrag}, the temperature distribution is relatively round and smooth. The only significant deviation from sphericity is in the protostellar disk. Thus, the gas around the protostar is heated quite uniformly, and outside the disk, where shear suppresses most fragmentation, there are no cold spots favorable to fragmentation. This is to be expected: the entire core is very optically thick, so radiation diffuses outward rather than free-streaming. As a result, there is very little shadowing, and clumps that are only starting to collapse are not sufficiently overdense to exclude the radiation field and remain cooler than their surroundings. As I discuss in \S~\ref{feedback}, only when gas reaches the densities typical of accretion disks or when optically thick structures begin to form can there can be significant temperature anisotropies due to collimation.

\begin{figure}
\begin{center}
\includegraphics{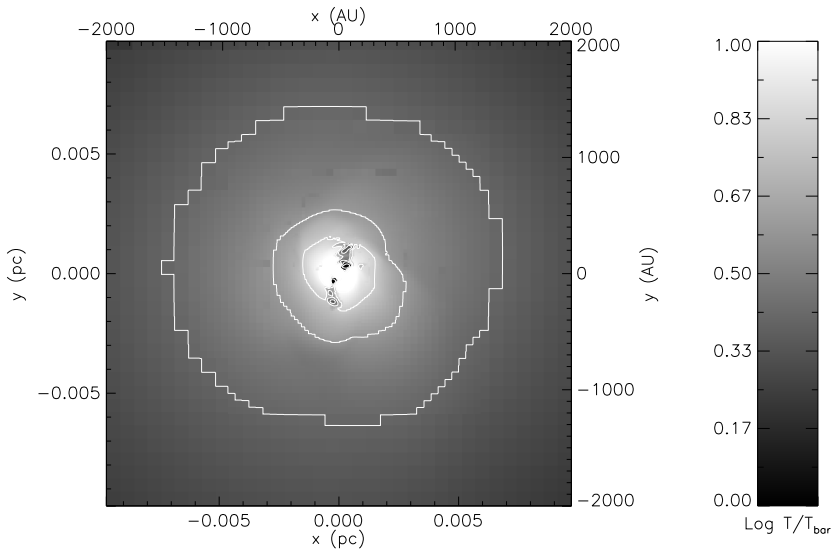}
\end{center}
\caption{
\label{tratiofrag}
Ratio of the temperature distribution shown in Figure \ref{tempfrag} to the temperature one would infer using the barotropic equation of state of \citet{dobbs05}. Note that the color scale starts at a ratio of 1.0, so any color other than black indicates that the true temperature is higher than the barotropic temperature. The outermost contour corresponds to a ratio of 0.4 dex, increasing by 0.2 dex with each subsequent contour.
}
\end{figure}

The weak fragmentation shown in simulations with radiation is strikingly different from what one obtains without it, where the number of fragments generally approaches the number of thermal Jeans masses in the initial cloud. Figure \ref{tratiofrag} shows why: the barotropic equation of state severely underestimates the temperature, making fragmentation far easier than it should be. The magnitude of the underestimate ranges from factors of a few at distances of thousands of AU to orders of magnitude in the central hundreds of AU. Since the Jeans mass depends on temperature to the 1.5 power, the error in the critical mass for fragment growth is larger still. It is easy to understand intuitively why the barotropic approximation fails so badly: the physical assumption underlying the barotropic approximation is that above some critical density the gas cannot radiate efficiently, and all its gravitational potential energy is converted into heat. However, 3D simulations of star formation cannot resolve stellar surfaces, so any gas that falls into sink particles of radius $\sim 10$ AU disappears from the simulation, taking its gravitational potential energy with it. However, since potential energy varies as $r^{-1}$, the vast majority of the energy is released in the final plunge from $\sim 10$ AU to the stellar surface. In the barotropic approximation one simply ignores this energy, which is the dominant source of heating until nuclear burning begins.

\section{Competitive Accretion}
\label{compaccretion}

Weak fragmentation means that a significant fraction of the mass in a massive core will end up in a single star or a few stars. However, for cores to be the direct progenitors of massive stars, it must also be the case that any additional mass a star accretes from outside its parent core is small compared to the stellar mass. The idea that most of a star's mass comes not from a parent core, but from gas in the cluster-forming clump that was not originally bound to that star, is called competitive accretion \citep[and references therein]{bonnell06a}. A number of numerical simulations appear to show exactly this process, and several authors have made simple theoretical models based on these. In these models, all stars are born from cores at roughly equal masses, with the initial mass ranging from brown dwarf masses \citep{bate05} to as much as $\sim 0.5$ $\msun$, the peak of the IMF, in the most recent models \citep{bonnell06c}. In these most recent models, most of these seeds do not accrete much mass in addition to that in their parent core, but a few stars are in special locations that allow them to undergo rapid accretion, reaching high masses. This process determines the IMF above the peak.

\subsection{Under What Conditions Does Competitive Accretion Occur?}

Competitive accretion definitely occurs in some simulations, and there is no reason to doubt that those simulations produce the correct result for the physics they include. However, in order to determine whether the simulations accurately reflect the properties of real star-formng clumps, it is necessary to investigate the physics behind the competitive accretion process. \citet{krumholz05e} defines the fractional mass change $f_M \equiv \dot{M}_* t_{\rm dyn}/M_*$ as the fractional mass change that a star of mass $M_*$ undergoes per dynamical (crossing) time $t_{\rm dyn}$ of its parent clump, where $\dot{M}_*$ is understood to refer to the accretion rate after the star has consumed its initial bound core. Competitive accretion models require $f_M \gg 1$. Accretion of gas that is not initially bound to a star can occur in one of two forms: either the star may capture other gravitationally bound cores and then accrete them, as proposed for example by \citet{stahler00}, or it may accrete gas that is not organized into bound structures.

The former process is reasonably easy to understand, since it is simply an extension of standard calculations of collision rates in stellar dynamics. The only significant complication is that collisions between stars and cores that occur at relative velocities that are too large do not result in capture, since the star will simply plough through the core without dissipating enough energy for the two to become bound. Even with this complication, the calculation is relatively straightforward, and \citet{krumholz05e} show that the fractional mass change due to captures of cores with mass comparable to the stellar mass $M_*$ in a star-forming clump of mass $M$ is
\begin{equation}
f_{M-\textrm{cap}} \approx 0.4 \phi_{\rm co} \left[4 + 2u^2 - (4 + 7.32 u^2) \exp(-1.33 u^2)\right],
\end{equation}
where $\phi_{\rm co}$ is the fraction of the clump mass that is in bound cores, $u \approx 10 \avir^{-1} (M_*/M)^{1/2}$ is the ratio of the escape velocity from the surface of a core to the velocity dispersion in the clump, and $\avir$ is the virial parameter for the core, roughly its ratio of turbulent kinetic energy to its gravitational potential energy. The significant thing to notice about this expression is that it does not approach unity unless $u$ is quite large, which in turn only happens for virial parameters $\avir \ll 1$, i.e. for clumps where the turbulent velocity dispersion is small compared to what is needed to prevent collapse.

\begin{figure}
\begin{center}
\includegraphics{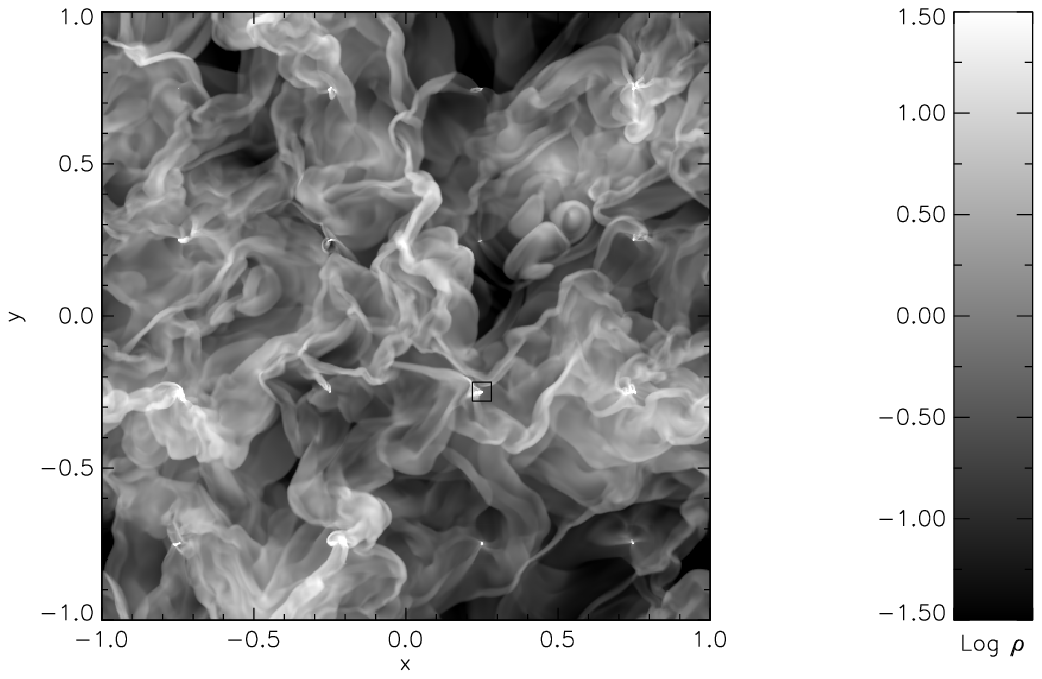}
\includegraphics{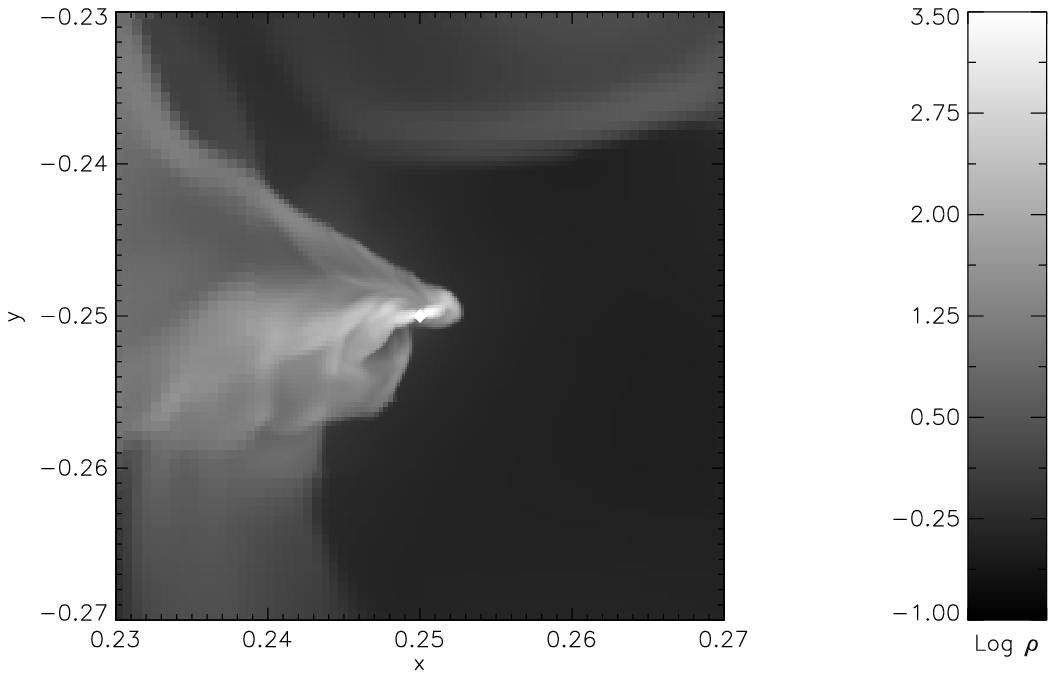}
\end{center}
\caption{
\label{bhturbimage}
Slices through a simulation of Bondi-Hoyle accretion in a turbulent medium by \citet{krumholz06a}, showing density in the entire simulation box (\textit{upper panel}) and in a small region (indicated by the black box) around one of the accreting particles \textit{(lower panel)}. The position of the particle is indicated by the small white diamond. The density and length are in dimensionless units where the mean density in the box is unity, and the box extends from $-1$ to $1$. The maximum resolution of the simulation is $8192^3$.
}
\end{figure}

\begin{figure}
\begin{center}
\includegraphics{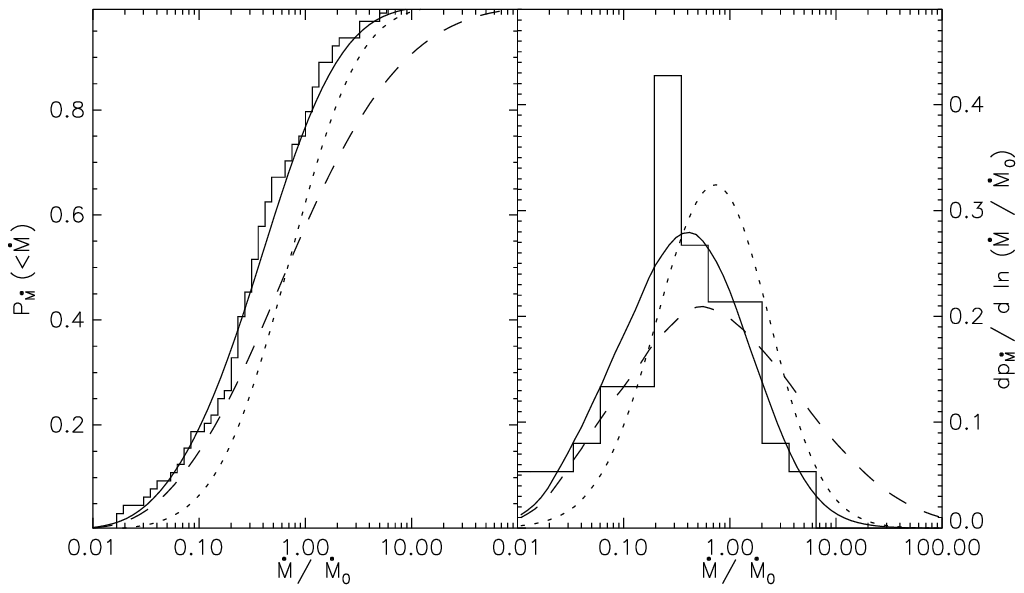}
\end{center}
\caption{
\label{bhturbpdf}
Cumulative (\textit{left panel}) and differential (\textit{right panel}) distribution of accretion rates  measured for particles in the simulation shown in Figure \ref{bhturbimage}. Accretion rates are normalized to $\dot{M}_0 \equiv 4\pi \overline{\rho} (G M_*)^2/(\sqrt{3} \sigma)^3$, where $\overline{\rho}$ is the mean density, $M_*$ is the stellar mass, and $\sigma$ is the 1D velocity dispersion in the simulation box. The histogram shows the simulation results, the solid line shows the \citet{krumholz06a} model, and the dashed and dotted lines show alternative models.
}
\end{figure}

Accretion of unbound gas is somewhat more complex, since to determine the accretion rate one requires a theoretical model for Bondi-Hoyle accretion in a turbulent medium. \citet{krumholz05b, krumholz06a} have developed such a theory and shown that it reproduces the results of simulations quite well. Figure \ref{bhturbimage} shows a sample of an adaptive mesh refinement simulation in which a grid of 64 Eulerian sink particles \citep{krumholz04} are placed into a turbulent medium and allowed to accrete until the mean accretion rate reaches equilibrium. Figure \ref{bhturbpdf} shows a comparison of the model prediction for the probability distribution of accretion rates to the simulation results. The model predicts, and the simulation confirms, that the mean accretion rate for a star of mass $M_*$ accreting from a medium of mean density $\overline{\rho}$ and 1D velocity dispersion $\sigma$ is
\begin{equation}
\dot{M}_* 4\pi \phi_{\rm BH} \approx \overline{\rho}\frac{(GM_*)^2}{(\sqrt{3}\sigma)^3},
\end{equation}
where the quantity $\phi_{BH}$ is a function of the Mach number and size scale of the turbulent region, an approximate analytic form for which is given in \citet{krumholz06a}. For the properties of observed star-forming regions, it is generally $\la 5$. From this result, one can compute $f_M$ due to accretion of unbound gas in a star-forming clump of mass $M$:
\begin{equation}
f_{M-\textrm{BH}} \approx 10\phi_{\rm BH} \avir^{-2} (M_*/M).
\end{equation}
Again, for cluster-clumps hundreds to thousands of $\msun$ in mass, $f_{M-\textrm{BH}}$ can be of order unity only for $\avir \ll 1$.

Combining the two potential sources of mass, one can derive an approximate criterion that a star-forming gas clump must satisfy in order for competitive accretion to occur within it. For seed stars of mass 0.5 $\msun$, this condition is
\begin{equation}
\label{cacondition}
\avir^2 M \la 50\,\msun.
\end{equation}
Straightforward application to observed star-forming clumps shows that they are nowhere near meeting this condition, since their typical masses are many hundreds to thousands of $\msun$, and their observed virial parameters are generally near unity. From this, \citet{krumholz05e} conclude that competitive accretion does not occur in real clumps. It occurs in simulations only because in those simulations there either was very little turbulence present in the initial conditions \citep[e.g.][]{klessen00a,klessen01}, or the initial turbulence has decayed away \citep[e.g.][]{bonnell03}, leaving the clumps sub-virial.

In light of this claim, \citet{bonnell06c} have re-analyzed the simulations of \citet{bonnell03}. They argue that competitive accretion occurs for a few stars that sit at the center of collapsing regions. These regions are stagnation points of the larger turbulent flow, where the velocity dispersion is much smaller than the mean and the density much larger. Thus, although the clump as a whole is turbulent, the regions where massive stars form are effectively decoupled from the large-scale flow. Since these decoupled regions have masses much smaller than that of the entire clump, and their virial parameters are smaller than unity due to their state of global collapse, they satisfy the competitive accretion condition (\ref{cacondition}), and stars within them gain mass via competitive accretion. 

I discuss whether such decoupled regions exist in real clumps below, in \S~\ref{globalcollapse}. However, one potential difficulty with the idea that competitive accretion occurs in locally detached parts of the flow is that it assumes that such detached regions would fragment and still produce many stars to compete, as it does in non-radiative simulations. However, these collapsing regions bear a striking resemblance to the objects we observe as massive cores: they are bound, high density regions with coherent velocity structures. As shown is \S~\ref{fragmentation}, such objects do not fragment strongly when one includes radiative transfer. Indeed, \citet{bate05} estimate the typical size scale of these collapsing regions as $\sim 1000$ AU, well inside the effective heating radius of a single rapidly accreting star. Instead of competitive accretion, the true behavior might simply be monolithic collapse of a massive core to a single system. In this case competitive accretion would be nothing more than core accretion simulated without a sufficiently accurate treatment of radiative transfer. Investigating that possibility will have to wait for future simulations. 

\subsection{Competitive Accretion and Global Collapse}
\label{globalcollapse}

The \citet{bonnell06c} analysis of competitive accretion concurs with \citet{krumholz05e} that competitive accretion cannot occur under typical conditions in a star forming clump. Instead, the possibility of competitive accretion depends on the existence of long-lasting stagnation points within which the gas and stars move together with gravity as the only significant force. The existence or non-existence of such stagnation points is something that is at least indirectly subject to observational determination. As \citet{bonnell06c} point out, if any source adds a significant amount of turbulent kinetic energy to the star-forming clump, such that the rate of energy injection is comparable to the rate at which energy is lost through radiative shocks, then ram pressure from flowing gas will push on gas but not on stars. No long-lived stagnation points will exist, and gas and star velocities will become decoupled. In this case, stars will randomly sample the density and velocity field of the clump, consistent with the treatment of \citet{krumholz05e} and \citet{krumholz06a}, and competitive accretion will not occur.

The most likely candidate for significant energy injection on the size scales of cluster-forming molecular clumps is feedback from protostellar outflows \citep{norman80}. Observational efforts to estimate whether the kinetic energy added by outflows is significant tentatively conclude that it is \citep{williams03, quillen05}. While these results are preliminary, if they are confirmed then competitive accretion cannot occur. One can also look for signs of large-scale collapse onto stagnation points. While there are a few examples of apparent infall signatures \citep{motte05, peretto06} on the scale of thousand $\msun$ star-forming clumps (as opposed to onto individual protostars, which is expected whether competitive accretion occurs or not), the majority of searches for infall signatures have turned up negative \citep{garay05}. Competitive accretion predicts that infall should be ubiquitous. It is, however, possible that these non-detections are due to observational confusion or lack of resolution.

A potentially more promising approach is to look for indirect signs of significant energy injection from feedback. If not, then the turbulence in protocluster gas clumps should decay rapidly \citep[e.g.][]{stone98}, leading to global collapse in roughly a crossing time. Over this time, a substantial fraction of the gas should be converted into stars. We can therefore test whether competitive accretion can occur by observationally estimating the time scale of star cluster formation and the star formation rate in pre-cluster gas.

\begin{figure}
\begin{center}
\includegraphics[scale=0.5]{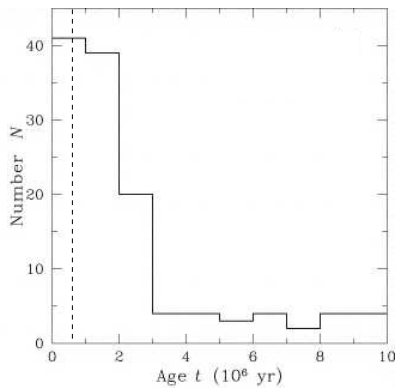}
\end{center}
\caption{
\label{ic348age}
Number of stars out of a sample of 162 versus age in IC348 (\textit{histogram}), and dynamical time in the cluster (\textit{dashed vertical line}). The histogram is reprinted from \citet{palla00} with permission from the authors.
}
\end{figure}

On the question of time scale, \citet{elmegreen00} has argued that star formation proceeds in a crossing time, but a more careful examination of much of the same data by \citet{tan06a} shows that the typical time scale is closer to $3-5$ crossing times ($6-10$ free-fall times). There are several lines of evidence for this, including the fairly round morphologies of most protocluster clumps, the lack of strong sub-clustering apparent in young clusters such as the Orion Nebula Cluster, the estimated momentum carried by the combined protostellar outflows of forming clusters, the age spreads of stars determined by fitting to pre-main sequence evolutionary tracks, and the ages of dynamical ejection events. In contrast, in the simulation of \citet{bonnell03} where feedback is neglected, star formation has completely ended and more than 50\% of the gas has been accreted in 1 crossing time, $\la 5\times 10^5$ yr. Figure \ref{ic348age} shows an example of how this prediction for the time scale of star cluster formation compares to observations: the solid line shows the observed ages of stars in IC348, a cluster whose most massive member is a mid-B star, computed by \citet{palla00}. The dashed vertical line shows the crossing time of $\approx 0.6$ Myr, computed by \citet{tan06a}. If the cluster formed in global collapse, all the stars should be to the left of the line. In this case and all the others analyzed by \citet{tan06a}, the observed age spreads and formation times of clusters are strongly incompatible with the idea that star-forming clumps are in a state of global collapse, and hence are strongly incompatible with competitive accretion.

\begin{figure}
\begin{center}
\includegraphics{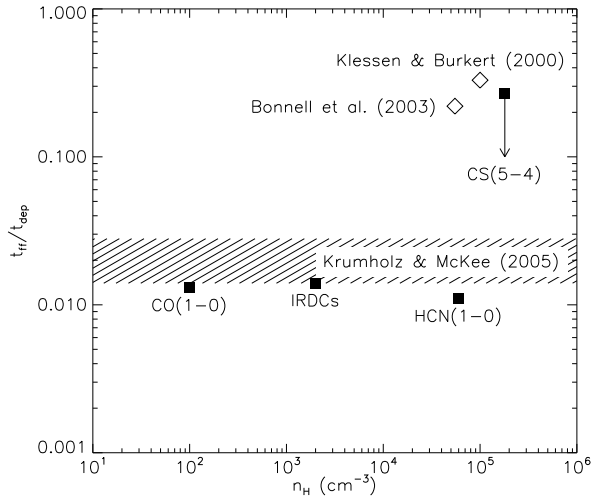}
\end{center}
\caption{
\label{sfdensefig}
Ratio of the free-fall time to the depletion time (defined as the time required to convert all gas into stars at the current star formation rate), estimated for gas at a given characteristic density. The plot shows observationally-determined points (\textit{solid squares}) for gas traced by CO(1-0) emission (giant molecular clouds), infrared dark clouds taken from the \citet{simon06} catalog, gas traced by HCN(1-0) emission \citep{gao04a,wu05}, and gas traced by CS(5-4) emission \citep{shirley03}. The CS(5-4) point is an upper limit. The plot also shows points from simulations (\textit{open diamonds}) by \citet{klessen00a} and \citet{bonnell03}, and the \citet{krumholz05c} model (\textit{hatched region}) applied under the assumption that star formation occurs in clouds with virial parameters $\avir=1-2$ and 1D Mach numbers $\calm=20-40$. The plot is adapted from \citet{krumholz06c}.
}
\end{figure}

One can also examine the star formation rate and compare to populations of dense gas clumps, as in the recent analysis by \citet{krumholz06c}. As an example, \citet{gao04a} and \citet{wu05} have shown that there is a strong correlation between luminosity in the HCN(1-0) line and infrared luminosity that extends from individual cluster-forming gas clumps in the Milky Way up to entire ultraluminous infrared galaxies. Since the infrared emission traces the star formation rate \citep[e.g.][]{kennicutt98b}, and HCN(1-0) emission traces the gas mass at densities $n \approx 6 \times 10^4$ H cm$^{-3}$ \citep{gao04b}, a typical density for protocluster gas, this correlation is a direct measure of the star formation rate in protocluster gas. \citet{wu05}, based on observationally-calibrated conversions from IR light to star formation rate and HCN(1-0) emission to gas mass find that their correlation corresponds to a star formation law $\dot{M}_* \approx M_{\rm HCN}/(80\mbox{ Myr})$, where $M_{\rm HCN}$ is the mass of gas emitting HCN(1-0). Including some corrections to those calibrations, \citet{tan06a} estimate a star formation law
\begin{equation}
\dot{M}_* \approx \frac{M_{\rm HCN}}{17\mbox{ Myr}}.
\end{equation}
We can compare this observational law to the predictions of theoretical models. \citet{bonnell03} simulate a gas clump with a mean density of $n=5.5\times 10^4$ H cm$^{-3}$, almost exactly the same as the density of observed HCN clumps. They find that clusters form in a free-fall collapse in which 58\% of the gas is incorporated into stars after 2.6 free-fall times. At the HCN density of $6\times 10^{4}$ H cm$^{-3}$, the free-fall time is 0.18 Myr, so the star formation law one would infer from the simulations is
\begin{equation}
\dot{M}_* \approx \frac{M_{\rm HCN}}{0.8\mbox{ Myr}},
\end{equation}
inconsistent with the observed relation. In contrast, if one assumes that HCN-emitting gas clumps are  virialized, $\avir \sim 1.5$, and turbulent, Mach number $\calm \sim 25$, then the \citet{krumholz05c} estimate for the star formation rate predicts
\begin{equation}
\dot{M}_* \approx \frac{M_{\rm HCN}}{10\mbox{ Myr}}.
\end{equation}
This is in good agreement with observation, and suggests that HCN-emitting gas cannot be in a state of global collapse, either in the Milky Way or in other galaxies. \citet{krumholz06c} find that repeating this exercise for other populations of star-forming objects gives similar results, as shown in Figure \ref{sfdensefig}.

In addition to the observational data, simulations of star cluster formation including feedback are beginning to appear in the literature, and these also cast doubt on the idea of global collapse or competitive accretion. \citet{li06b} simulate a star-forming clump with a code that includes an approximate treatment of the energy and momentum injected by outflows from newly formed stars. In their simulations, there are no long-lived stagnation points of the type required for competitive accretion. Any such stagnation points are rapidly disrupted either by internal feedback from stars within them or by external shocking. Energy injected by feedback replenishes energy lost in radiative shocks, so the cloud remains virialized and does not collapse. Instead, it approaches a roughly constant velocity dispersion. In contrast to the results of \citet{bonnell03} without feedback, \citet{li06b} find that only $\sim 6\%$ of the mass collapses into stars per free-fall time, in reasonable agreement with both observational estimates of the cluster formation time scale and the star formation rate in dense gas.

In summary, the question of whether competitive accretion occurs reduces to the question of whether cluster-forming gas clumps are in a state of global collapse, and both observational estimates and simulations including feedback seem to rule out this possibility. The mass in a protostellar core is all the mass that a star will ever have.

\section{Feedback and Accretion}
\label{feedback}

As shown in the previous sections, massive cores do not fragment strongly, nor do the stars that form within them gain significant mass from outside their parent core. There remains, however, the problem of getting mass from the core onto a star. This is potentially difficult because massive protostars have short Kelvin-Helmholtz times that enable them to reach the main sequence while they are still forming from their parent clouds \citep{shu87}. Once nuclear burning begins, the star will have the immense luminosity of a main sequence star of the same mass. Radiation might inhibit accretion in two different ways: first, it exerts a direct radiation pressure force on dust grains suspended in the incoming gas, and this force could be stronger than gravity. Second, ionizing photons can create an HII region around the star, within which the temperature is $10^4$ K and the sound speed is 10 km s$^{-1}$. The HII region is at very high pressure, and may therefore be able to expand and unbind the accreting gas. If either of these feedback processes is capable of halting accretion, then the mapping from core mass to star mass will not be simple at the high mass end, and it will not be possible to explain the properties of stars in terms of the properties of cores.

\subsection{Radiation Pressure Feedback}

Early spherically symmetric calculations of the effects of radiation pressure on accretion flows onto massive stars found that radiation becomes stronger than gravity, and halts accretion, at masses of $20-40$ $\msun$ \citep{kahn74, wolfire87} for typical Galactic metallicities. More recent work has relaxed this limit by considering the effects of accretion disks, which allow most of the mass to accrete from within the optically thick disk, where it is shielded from radiation pressure \citep{nakano89,nakano95,jijina96}. Furthermore, rotational flattening can cause the radiation field itself to become asymmetric, as more radiation escapes in the lower density polar direction than through the higher density equatorial plane. Cylindrically-symmetric radiation-hydrodynamic simulations show that this flashlight effect allows accretion to slightly more than 40 $\msun$ \citep{yorke02}. These simulations also reveal that forming a massive star is easier in simulations that include a multifrequency treatment of the radiation field than in those that use a gray approximation, probably because the collimation of the radiation field is reduced by the gray approximation.

\begin{figure}
\begin{center}
\includegraphics[scale=0.8]{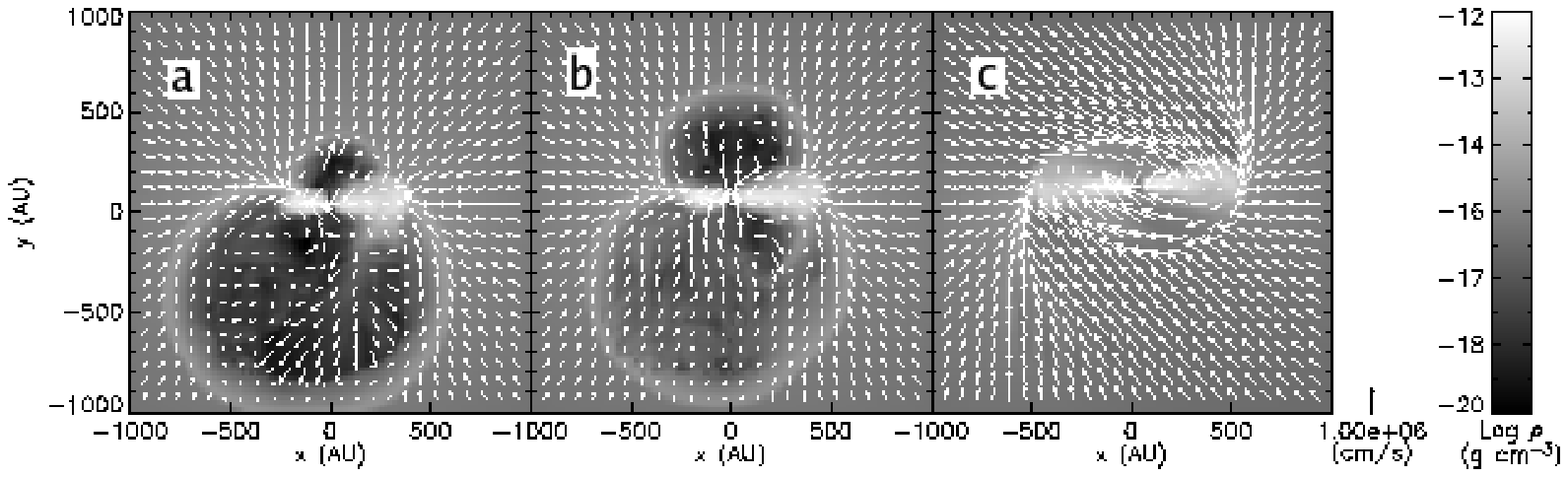}
\end{center}
\caption{
\label{hmbubble}
Slices in the XZ plane showing the density (\textit{grayscale}) and velocity (\textit{arrows}) in a simulation of the collapse of a 100 $\msun$ core by \citet{krumholz05d}. The panels show times of (a) $1.5\times 10^4$ yr, (b), $1.65\times 10^4$ yr, and (c) $2.0\times 10^4$ yr after the start of the simulation. The corresponding stellar masses at those times are $21.3$ $\msun$, $22.4$ $\msun$, and $25.7$ $\msun$. Note that these masses are considerably larger than those of the turbulent runs discussed in \S~\ref{fragmentation} at the same times primarily due to the steeper initial density profile assumed in these simulations.
}
\end{figure}

\begin{figure}
\begin{center}
\includegraphics[scale=0.45]{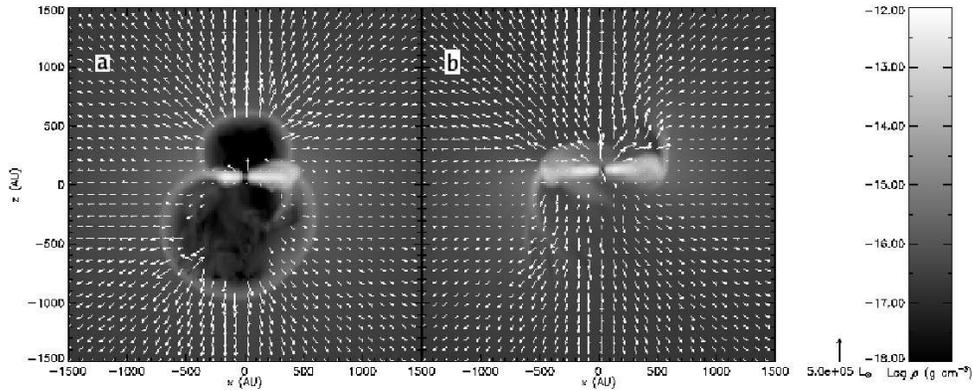}
\end{center}
\caption{
\label{hmflux}
Slices in the XZ plane showing the density (\textit{grayscale}) and radiation flux multiplied by $4\pi r^2$ (\textit{arrows}) in the simulation shown in Figure \ref{hmbubble}. For clarity, the flux vectors in the optically thin bubble interior in panel (a) have been omitted. The times shown in panels (a) and (b) are the same as those shown in panels (b) and (c) of Figure \ref{hmbubble}.
}
\end{figure}

More recent 3D simulations show that the flashlight effect persists into 3D, and is actually enhanced by a qualitatively new effect. Figure \ref{hmbubble} shows time slices from a radiation-hydrodynamic simulation of the collapse of a 100 $\msun$, 0.1 pc-radius core with an initial density profile $\rho\propto r^{-2}$, performed by \citet[and 2006, in preparation]{krumholz05d}. The initial core has a temperature of 40 K and no turbulence, just overall solid body rotation at a rate such that the rotational kinetic energy is 2\% of the gravitational potential energy, a typical rotation rate observed for low mass cores \citep{goodman93}. The simulation shows that, at stellar masses $\la 17$ $\msun$, stellar radiation is too weak to reverse inflow, and the gas falls onto an accretion disk and accretes in a steady, cylindrically-symmetric flow. At about $17$ $\msun$, stellar radiation begins to reverse inflow and drive radiation bubbles above and below the accretion disk. The interiors of these bubbles are optically thin and very low density, while the walls reach densities $\ga 10^{10}$ H cm$^{-3}$. Gas reaching the bubbles ceases to move radially inward. Instead, it travels along the bubble wall until it falls onto the accretion disk. At that point, it is shield from stellar radiation and is able to accrete onto the star. As shown in Figure \ref{hmflux}, the bubbles also collimate radiation, increasing the flux in the polar direction and decreasing it in the equatorial plane. At the time shown in Figure \ref{hmflux}a, the radiation flux just outside the radiation bubble in the polar direction exceeds that in the equatorial plane at the same radius by more than an order of magnitude.

The bubbles expand asymmetrically with respect to both the polar axis and the equatorial plane due to an instability. The direction in which a bubble has expanded farthest and cleared out a low optical depth region represents the path of least resistance for radiation leaving the star to escape to infinity. As a result, radiation is collimated in whichever direction the bubble has expanded farthest. In turn, this causes the radiation pressure force to be largest in that direction, which amplifies the rate of bubble expansion. Figure \ref{hmbubble}a shows this asymmetric expansion phase. As the simulation progresses the instability becomes more violent and the bubbles start to collapse, as shown in Figure \ref{hmbubble}b. The collapse is due to radiation Rayleigh-Taylor instability: the gas is a heavy fluid, and the radiation is a light fluid trying to hold it up, an unstable configuration. Once the bubbles collapse, as shown in Figure \ref{hmbubble}c, infalling gas is deflected by radiation onto the dense remnants of the bubble walls, which are optically very thick and therefore self-shielding against radiation. The gas travels along the walls onto the disk and then accretes. In this configuration the remnant walls collimate the accretion, as shown in Figure \ref{hmflux}b. The radiation flux behind the walls is very small, so gas accreting from that direction feels very little radiation pressure force.

Simulations to date have reached $\approx 34$ $\msun$ in this manner, with $5-10$ $\msun$ more in the accretion disk and no sign of a reversal of infall. Note that these calculations use gray radiative transfer, for which case \citet{yorke02} found an upper limit of about 20 $\msun$ for the stellar mass. This suggests that multifrequency 3D calculations would show even easier accretion.

\citet{krumholz05a} point out a second mechanism for overcoming radiation pressure feedback. Massive protostars have outflows that appear to be simply scaled-up versions of the outflows from lower mass stars (see \citealt{beuther05} for a recent review), or that are somewhat wider in angle for the most massive, O type stars. These outflows can have a very strong effect on the radiation field. Outflows are launched from the inner accretion disk around the star, where the gas is so hot that any dust grains present within it sublime. As a result, outflows contain almost no dust when they are launched, and simple calculations show that due to the velocities of $\ga 500$ km s$^{-1}$ with which they flow, there is insufficient time for large dust grains to reform before the outflowing gas reaches distances $\ga 0.1$ pc from the source star. This makes the outflow cavities optically thin, so they can collimate radiation and carry it out of the optically thick accreting envelope very effectively. \citet{krumholz05a} use Monte Carlo radiative transfer calculations to explore this effect, and show that it can reduce the radiation pressure force on infalling gas by an order of magnitude. This can in turn shift the radiation pressure force from a regime where it is stronger than the gravitational force to one where it is weaker than gravity, allowing accretion to continue where it might have been halted without the protostellar outflow.

It is unclear how this effect will interact with radiation pressure bubbles, since the simulations in which the bubbles appear do not yet include a model for protostellar outflows. Nonetheless, the overall conclusion one may draw from the two effects is that radiation pressure cannot easily halt accretion. There are at least two mechanisms which are capable of rendering the radiation field aspherical, allowing accretion over the range of solid angle where the radiation pressure force is reduced. Gravity is very effective at channeling gas into those solid angles, and then onto the star. 

\subsection{Ionization}

The second form of radiative feedback that accretion onto a massive star must overcome is ionization, first recognized as a problem by \citet{larson71} and \citet{yorke77}. If the ionizing photons from a massive protostar are able to reach gas at distance from the star where the escape speed is $\la 10$ km s$^{-1}$ then the ionized gas will be unbound and may escape to infinity. However, \citet{walmsley95} shows that sufficiently rapid accretion can avoid this problem by keeping the ionized region quenched against the stellar surface. For spherically symmetric infall, if the accretion rate is larger than
\begin{equation}
\dot{M}_{\rm crit} = \sqrt{\frac{4\pi G M_* S m_H^2}{\alpha^{(2)}}}
\approx 2\times 10^{-5}
\left(\frac{M_*}{10\,\msun}\right)^{1/2} \left(\frac{S}{10^{49}\mbox{ s}^{-1}}\right)^{1/2}\;\msun\mbox{ yr}^{-1}, 
\end{equation}
where $S$ is the ionizing luminosity of the star in photons s$^{-1}$ and $\alpha^{(2)}$ is the recombination coefficient to excited levels of hydrogen, then all ionizing photons will be absorbed near the stellar surface. The escape velocity there is much larger than 10 km s$^{-1}$, so this ionized gas will be confined by stellar gravity and accretion may continue unimpeded.

Observational estimates of HII region lifetimes indeed appear to require confinement of this sort. Ultracompact HII regions, roughly those $\la 0.1$ pc in radius, have a dynamical expansion time $\la 0.1\mbox{ pc}/10\mbox{ km s}^{-1} \approx 10^4$ yr, but the number of observed ultracompact HII regions implies that their lifetimes must be closer to $10^5$ yr \citep{wood89, kurtz94}. Moreover, some observed systems show signs of both accretion and the presence of an ultracompact HII region \citep[e.g.][]{hofner99, sollins05}, providing direct evidence that accretion can continue past HII region formation.

\citet{hoare06} gives a recent review of ultracompact HII regions, and \citet{keto06} attempts to construct a unified theoretical framework within which one can understand the relationship between HII regions and accretion \citep[though see][]{beuther05c}. In this model, the behavior of an HII region is controlled by the relative sizes of the ionization radius within which the ionization and recombination rates balance and all ionizing photons are absorbed, the gravitational radius where a spherical accretion flow accelerates from subsonic to supersonic relative to the sound speed in ionized gas, and the disk radius, at which angular momentum in accreting gas causes infall to become aspherical and a disk to form. This model can roughly explain observed HII region morphologies. One of its consequences is that if the ionization radius is smaller than the gravitational radius, the HII region will remain confined and accretion will continue unimpeded. For the accretion rates of $10^{-4}-10^{-3}$ $\msun$ yr$^{-1}$ one expects given the conditions in massive cores, massive protostars will remain in this regime until they have exhausted essentially all of their available gas. Thus, ionization feedback does not present a significant barrier to accretion.

\section{Conclusions and Future Directions}
\label{future}

\subsection{Massive Star Formation from Massive Cores}

The results summarized in the previous sections combine to give a unified scenario of massive star formation: massive stars form by the gravitational collapse of massive cores. We observe these cores, from a few tenths to a few hundred $\msun$ in mass, in the dense molecular clumps where star clusters form. Their mass distribution matches the stellar IMF, and their spatial distribution is roughly consistent with the spatial distribution of massive stars in clusters. When massive cores collapse, they do not fragment strongly because radiation feedback from the high accretion rates they produce warms their inner regions, raising the Jeans mass and suppressing fragmentation. As a result, all of their mass falls onto one a few stars. This occurs even though the cores are turbulent and contain complex density and velocity structures.

The stars that form in massive cores do not gain significant mass from outside their parent cores, because gas that is not gravitationally bound to them at birth is too turbulent for significant accretion to occur. The mostly likely agent for keeping the gas turbulent is feedback from protostellar outflows, and both observational estimates and simulations suggest that this mechanism works. Regardless of the agent, though, observations of the age spread of young clusters and the star formation rate in pre-cluster gas unambiguously require that star cluster form from gas that is not in free-fall collapse.

The gas falling onto a massive protostar at the center of a massive core must accrete in the face of significant radiation feedback in the form of radiation pressure on dust grains suspended in the inflowing gas, and ionizing radiation that generates high pressure ionized gas. However, neither mechanism substantially impedes accretion. Radiation escapes through low optical depth channels, which either form spontaneously due to radiation-hydrodynamic instabilities or are created by protostellar outflows. Gas accretes through high optical depth channels onto an accretion disk, where it is shielded from protostellar radiation. At the same time, the accretion flow is capable of absorbing all the ionizing stellar photons near the star, where the escape speed is larger than the ionized gas sound speed. As a result, formation of an HII region is either suppressed entirely, or the ionized region is kept close to the star where gravity confines it and prevents it from expanding. Accretion continues through the ionized region.

Because gas from a massive core accretes onto one or two stars, with no significant impedance from feedback, and those stars do not gain a significant amount of mass from outside their parent cores, there is a direct mapping from the properties of cores to the properties of stars. Consequently the mass and spatial distributions of young clusters are direct imprints of the mass and spatial distributions of cores in their gas phase progenitors, possibly with some additional redistribution of mass due to dynamical mass segregation over the several crossing times that it takes for a cluster to assemble. Observables in the gas phase connect directly to observables in the stellar phase.

\subsection{Problems for the Future}

While this scenario explains a great deal of what we observe about massive stars, it also omits a number of physical effects that a more complete picture should include, and it needs more rigorous testing against observation. In this section, I discuss the prospects for improvement in these areas in the next few years.

\subsubsection{Magnetic Fields}

None of the models of massive star formation proposed to date have included the effects of magnetic fields on the dynamics of the collapsing gas. Partly this is due to a lack of observational information about magnetic fields in massive star-forming regions. \citet{crutcher05} summarizes the state of observations, and concludes that magnetic fields are marginally dynamically significant in massive star-forming cores. However, this determination is extremely uncertain due to the difficulty of interpreting observational indicators of magnetic field strength. Part of this uncertainty comes from geometry. The critical magnetic flux required to hold up a core against gravity depends on the core shape and mass distribution, and differing assumptions about the shape can produce qualitatively different conclusions about how the magnetic field strength compares to that required for it to be dynamically significant \citep{bourke01}.

Beyond the geometric effect, measurements of field strengths themselves are quite uncertain. The two most commonly-used techniques for determining magnetic field strengths in molecular gas are Zeeman splitting of OH or CN lines, and the Chandrasekhar-Fermi method in which one uses the dispersion of polarization vectors seen in dust emission to estimate how the ratio of magnetic to kinetic energy density. Both of these techniques suffer from major systematic errors. Zeeman splitting averages the magnetic field along the line of sight, potentially washing out the signal, and is also only sensitive to regions where the observed species are present. Since many species freeze out onto dust grains in the densest parts of cores \citep{tafalla02}, Zeeman measurements may reveal more about magnetic field strengths in the diffuse gas around cores than in cores themselves. For Chandrasekhar-Fermi measurements, one also faces an uncertainty about where along the line of sight the observed polarized light is being emitted, and whether one is really measuring field in the core or in its outskirts. This uncertainty is made even worse by the fact that the magnetic field one deduces for a given dispersion of polarization vectors depends on the gas density, so uncertainty in the gas properties in the emitting region translates directly into uncertainty about the magnetic field strength.

The level of uncertainty is driven home by the result that, in at least some regions where both have been used, the inferred magnetic field strengths differ by a factor of $\sim 100$ \citep[e.g.][]{crutcher99, lai01, lai02}. While in principle this is possible because Zeeman splitting measures the field along the line of sight and Chandrasekhar-Fermi measures the field in the plane of the sky, it seems implausible that the magnetic field in multiple sources could be so perfectly aligned perpendicular to the line of sight. A more likely explanation is that there are large systematic errors in one or both techniques, or that the two techniques are simply measuring the field in different regions.

In the absence of more conclusive observational data, it is hard to know whether it is important to include magnetic fields in massive star formation models or not. In principle one could make models both with and without dynamically significant magnetic fields to explore their effects, but at present magnetohydrodynamics codes using either smoothed particle hydrodynamics or adaptive mesh refinement, one of which is required to achieve the requisite dynamic range to simulate star formation, are still in their infancy \citep[e.g.][]{price04a, crockett05}. There are no codes past the experimental stage that include both MHD and radiative transfer, which the discussion above demonstrates is absolutely crucial. The question of the influence of magnetic fields will therefore have to wait for progress in both code development and observations.

\subsubsection{Improved Simulation Physics}

In addition to magnetic fields, there are numerous other physical effects that the current generation of simulations ignore. It is chastening to realize that almost every new piece of physics that has been added to simulations of massive star formation has revealed a qualitatively new and unexpected phenomenon: adding radiative transfer to simulations of turbulent cores shows that fragmentation is much weaker than one might suppose from purely hydrodynamic calculations (\S~\ref{fragmentation}); moving from gray to multifrequency radiative transfer in two dimensional simulations doubles the maximum stellar mass formed due to enhancement of radiation beaming \citep{yorke02}; moving from two to three dimensions with gray radiative transfer reveals the formation of unstable, asymmetric radiation bubbles \citep[and \S~\ref{feedback}]{krumholz05d}; including protostellar outflows in models greatly reduces the efficacy of radiation feedback on the scale of individual cores \citep{krumholz05a}, and changes the dynamics of star cluster formation from rapid collapse and competitive accretion to quasi-equilibrium behavior where turbulent decay is balanced by injection of kinetic energy by outflows \citep{li06b}. This pattern suggests that there may be new effects yet to be discovered as simulation physics improves. An important corollary of this is that we have very likely reached the limit of what pure gravity plus hydrodynamics simulations with no feedback can teach us. Major work in the future will not come from simply running more and bigger simulations of the same type, but from adding new physics to the problem.

The next logical steps in modeling massive star formation include extending multifrequency radiative transfer to three dimensions, using radiation hydrodynamic codes to simulate the formation of star clusters instead of just individual stars, and including protostellar outflows in radiation-hydrodynamic simulations of massive cores. These are all relatively straightforward extensions of existing techniques. The primary limit to using them is computer power, since gray radiation-hydrodynamic simulations in three dimensions take months on available supercomputers. However, Moore's Law will help in this regard, and these simulations should become feasible in the next few years. More difficult are the problems for which not only computer power but also our current simulation methods are inadequate, and for which we will have to develop new techniques. One problem in this category is putting more realistic dust physics into three dimensional models, including coagulation, shattering, and differential motion of grains and gas, which may substantially change opacities \citep{preibisch95, sonnhalter95, suttner99}. Another is radiative transfer that goes beyond flux-limited diffusion \citep[e.g][]{hayes03}. Improvements in these areas are likely to take somewhat longer, perhaps becoming reasonable on the time scale of five or more years.

\subsubsection{Connecting Models to Observations}

A third area in which substantial progress must be made is connecting simulation and theoretical models to observations. There has been a fair amount of work on relating simulation results to statistical diagnostics of young clusters such as the IMF and binary and brown dwarf properties. However, these indicators are quite difficult to use as a means of distinguishing theories because they focus on the outcome only after the star cluster is fully formed, and there appear to be multiple mechanisms capable of producing the same outcome (e.g. \citealt{padoan02} and \citealt{bonnell03} on the origin of the IMF, or \citealt{bonnell05} and \citealt{krumholz06d} on the binarity properties of massive stars). In addition, observations of the results of star formation are difficult to connect to models because we do not know the properties of the progenitor gas system for a given stellar system. It is often possible to reproduce a given stellar system simply by tuning the properties of the assumed progenitor.

Better diagnostics are likely to come from comparing the results of simulations and theoretical models to observations of systems that are still actively forming stars. Masers provide one potential connection, since they make it possible to trace gas properties and kinematics on very small scales (e.g. \citealt{torrelles01}, who find expanding bubbles traced by masers, and \citealt{greenhill04} who detect a disk and an outflow cavity traced by masers). These may be particularly useful for tracing radiation-driven bubbles, since the density of $10^9-10^{10}$ cm$^{-3}$ and temperature of a few hundred K of bubble walls is favorable for maser emission. As discussed elsewhere in this volume, observational searches for accretion disks around massive stars, which are expected for massive star formation by core collapse but probably not for collisional formation, are another example. Morphologies of warm gas, as seen in high resolution observations of ``hot cores" \citep[e.g.][]{garay05}, provide another potential point of comparison to simulations. One can carry such comparisons even further by post-processing the simulation results with radiative transfer codes to produce simulated line profiles and intensity maps. One can also look for telltale signs of star formation by collisions, such as infrared flares \citep{bally05} or embedded clusters in which the stellar density reaches the $\sim 10^7$ pc$^{-3}$ required for efficient collisions -- such high densities should produce distinctive spectral energy distributions that peak at wavelengths approaching 100 $\mu$m \citep[and Chakrabarti \& McKee, 2006, in preparation]{chakrabarti05}. Yet another technique is to use extragalactic observations, in which one can observe the entire galactic population of a given sort of object \citep[e.g. HCN-emitting clouds][]{gao04a}, and then make arguments based on population statistics \citep[e.g.][]{krumholz06c}. Any or all off these are likely to be far more definitive than comparisons to stellar systems.

Prospects for improvement in this area of massive star formation theory are quite good. Current and planned millimeter instruments such as PdBI, the SMA, CARMA, and ALMA will allow higher resolution examination of more distant, embedded star-forming systems than has heretofore been possible, and surveys or both Galactic and extra-Galactic star-forming regions with Spitzer are providing large databases that we can use to test models on the level of populations. These sources of data have already begun to let us distinguish between models of massive star formation, and will provide even more powerful diagnostics in the next few years.

\acknowledgements I thank H. Beuther, R.~I. Klein, Z.-Y. Li, C.~F. McKee, F. Nakamura, J.~M. Stone, and J.~C. Tan for helpful discussions. This work was supported: NASA through Hubble Fellowship grant HSF-HF-01186 awarded by the Space Telescope Science Institute, which is operated by the Association of Universities for Research in Astronomy, Inc., for NASA, under contract NAS 5-26555; the Arctic Region Supercomputing Center; the National Energy Research Scientific Computing Center, which is supported by the Office of Science of the US Department of Energy under contract DE-AC03-76SF00098, through ERCAP grant 80325; the NSF San Diego Supercomputer Center through NPACI program grant UCB267; and the US Department of Energy at the Lawrence Livermore National Laboratory under contract W-7405-Eng-48.

%\bibliographystyle{cupconf}
%\bibliography{refs}

\end{document}